%% file: method.tex
\documentclass{article}
\title{Investigation of FWER and power for methodological changes introduced in the bdots R package}
\date{}
\author{Collin Nolte}
\usepackage{setspace}
\usepackage[margin=1in]{geometry}
\usepackage{amsmath}
\usepackage{graphicx}
\usepackage{datetime}
\usepackage{natbib}
\usepackage{float}
\newcommand{\xt}{\texttt}

\usepackage{subfigure}
\graphicspath{{img/}}
\usepackage{color}

\usepackage{lscape} 

\usepackage{algpseudocode}
\usepackage{algorithm}

\usepackage{listings}

\begin{document}

\maketitle

\begin{abstract}
In 2025, we identified a methodological issue in the bootstrapped differences of times series (BDOTS) first introduced in 2017 resulting in a significant inflation of the family-wise error rate. The goal of the present manuscript is threefold: to identify the problem in the original methodology, to present two alternative solutions, and to compare estimates of the FWER and power of each of the considered methods across a variety of experimental conditions. We find conclusive evidence that the original BDOTS method does inflate the FWER, while each of the proposed alternatives maintain a FWER much closer to the nominal rate. Additionally, we demonstrate the relationship between power and effect size for each of the proposed methods. In total, the results presented justify the methodological changes presented in the new iteration of the bdots package.
\end{abstract}

\section{Introduction}

A problem ubiquitous throughout the sciences, though in the cognitive sciences especially, is that of statistically analyzing a process unfolding in time. In particular, we consider the problem of comparing a process in time as it evolves differentially between two or more experimental groups. And while there are many techniques for demonstrating \textit{that} a difference exists, few offer any insight into \textit{when}. Testing for temporal differences is complicated by the fact that when this process is continuous, there are often an arbitrarily large number of time points that could be compared. This is essentially a problem of multiple comparisons.

One approach to this was introduced by \citet{oleson2017detecting} who presented a modified Bonferroni correction to a series of test statistics, using estimates of autocorrelation between statistics to make the appropriate adjustments to the significance levels to control FWER. This approach, which they named bootstrapped differences in time series, was introduced in the R package \xt{bdots} \citep{seedorff2018bdots}.

The modified Bonferonni correction used by \xt{bdots} relies on the construction of estimated distributions of time series via bootstrapping for each experimental group. These distributions are then used to construct test statistics at each observed time point. A closer look at the original iteration presents concerns as it involves restrictive assumptions on the data that are unlikely to be met in many, if not most situations. 

This includes data typically collected in the context of the Visual World Paradigm (VWP), a widely used experimental paradigm in language research that involves tracking eye movements from participants in response to spoken language. This is notable in that it was data collected in VWP research that motivated the \xt{bdots} methodology. There it was maintained that the process of interest in each experimental group assumed a homogeneous mean structure, with no between-subject variability to be accounted for, an assumption we will explore in more detail in the following section. Empirical data collected in a variety of contexts suggests that this assumption is unlikely to be true, the consequence of which is a family-wise error rate that is unacceptably high. 

This manuscript has three goals: to identify the issue in the original iteration of \xt{bdots} resulting in an inflated FWER, to propose two solutions to this problem, and to demonstrate the efficacy of the proposed solutions. We begin with a mathematical description of the problem, along with our proposed solutions. These consist of a modification to the original algorithm and the introduction of a permutation test. We then present a collection of simulations evaluating the FWER for each method across a variety of experimental conditions. We conclude with two simulations intended to demonstrate power among the competing methods. Ultimately, the results of these simulations are intended to justify the methodological changes that are now included in the current instantiation of the \xt{bdots} package \citep{bdots2025}.

\section{Methods}

In each of the methods to be described, we begin with the observation of $y_{it}$ for subjects $i = 1, \dots, n$ over times $t = 1, \dots, T$. Typically, these subjects fall into different groups $g = 1, \dots, G$, with each group containing $n_g$ subjects. We further assume that the empirically observed data derives its mean structure from a parametric function $f$ with associated error:
\begin{equation}\label{eq:mean_structure}
y_{it} = f(t|\theta_i) + \epsilon_{it}
\end{equation}
where $\theta_i$ is the subject-specific parametrization of $f$ with
\begin{equation}
\epsilon_{it} = \phi \epsilon_{i, t-1} + w_{it}, \quad w_{it} \sim N(0, \sigma).
\end{equation}

Under this paradigm, the errors are permitted to be either iid normal (with $\phi = 0$) or have an AR(1) structure, with $0 < \phi < 1$. It is generally assumed that the observed data across subjects make up a distribution for each group, specified as either a multivariate distribution of the parameters $\theta_i$ or a distribution of resulting curves, $f(t|\theta_i)$. We further assume that each subject's parameters are drawn from a group-level distribution,
\begin{equation}\label{eq:group_dist}
\theta_i \sim N(\mu_g, V_g).
\end{equation}
Ultimately, it will be from a distribution of curves that we determine the temporal characteristics of each group, with the differences in these temporal characteristics being what we are interested in identifying. With the general notation addressed, we now move to the particulars of each of the methods considered.

\subsection{Homogeneous Bootstrap}

The original bootstrapping algorithm presented in \citet{oleson2017detecting} follows what we will call the \textit{homogeneous means} assumption. Accordingly, we will call this bootstrapping algorithm the \textit{homogeneous bootstrap}. Under the homogeneous means assumption, it is still assumed that observed data for each subject retains the mean structure given in Equation~\ref{eq:mean_structure}, but with the additional assumption that $\theta_i = \theta_j$ for all subjects $i, j$ within the same group. In other words, there is assumed to be no variability in the mean structure between subjects within the same group. This is evidenced in the original bootstrapping differences in time series algorithm, which samples without replacement at each bootstrapping step:

\begin{enumerate}
\item For each subject, fit a nonlinear regression model to obtain $\hat{\theta}_i$. \citet{oleson2017detecting} recommends specifying an AR(1) autocorrelation structure for model errors. Assuming large sample normality, the sampling distribution of each estimator can be approximated by a multivariate normal distribution with mean for subject $i$, $\hat{\theta}_i$ corresponding to the point estimate and standard deviations corresponding to the standard errors, $s_i$.

\item Using the approximate sampling distributions in (1), randomly draw one bootstrap estimate for each of the model parameters on every subject 
\begin{equation}
\hat{\theta}_i^{(b)} \sim N( \hat{\theta}_i, s_i^2)
\end{equation}

\item Once a bootstrap estimate has been collected for each parameter and for every subject, for each parameter, find the mean of the bootstrap estimates across $n_g$ individuals for the $b$th bootstrap in group $g$,
\begin{equation}
\theta_g^{(b)} = \frac{1}{n_g} \sum_{i=1}^{n_g} \hat{\theta}_i^{(b)}
\end{equation}
\item Use the mean parameter estimates to determine a bootstrapped population level curve, which provides the average population response at each time point, $f(t| \theta_g^{(b)})$.

\item Perform steps (2)-(4) B times to obtain estimates of the population curves. Use these to create estimates of the mean response and standard deviation at each of the time points. For each group $g = 1, \dots, G$, this gives
\begin{equation}
\overline{p}_{gt} = \frac1B \sum_{b=1}^B f(t| \theta_g^{(b)}), \qquad s_{gt}^2 = \frac{1}{B-1} \sum_{b=1}^B  [f(t| \theta_g^{(b)}) - \overline{p}_{gt}]^2,
\end{equation}
where $\overline{p}_{gt}$ and $s_{gt}^2$ are mean and variance estimates at each time point for group $g$.
\end{enumerate}

Population means and standard deviations at each time point for each of the groups were used to construct a series of (correlated) test statistics, where the family-wise error rate was controlled by using the modified Bonferonni correction introduced in \citet{oleson2017detecting} to test for significance. As this correction is also used for the heterogeneous bootstrap presented next, we offer a more comprehensive review of this adjustment at the end of this section.

\subsection{Heterogeneous Bootstrap}

Typically, subjects within a group demonstrate considerable variability in their mean parameter estimates. In this case, we should avoid the presumption that $\theta_i = \theta_j$, as accounting for between-subject variability within a group will be critical for obtaining a reasonable distribution of the population curves. More likely, we may assume that the distribution of parameters for subjects $i = 1, \dots, n_g$ in group $g = 1, \dots, G$ follows the distribution
\begin{equation}\label{eq:theta_i_dist}
\theta_i \sim N(\mu_{g}, V_{g}),
\end{equation}
where $\mu_g$ and $V_g$ are the group-specific mean and variance values, respectively. In contrast to the previous set of assumptions, we call this the \emph{heterogeneous means} assumption Similar to what was presented in the homogeneous bootstrap algorithm, we can further account for uncertainty in our estimation of $\theta_i$ by $\hat{\theta}_i$ by treating the standard errors derived when fitting the observed data to the mean structure suggested in Equation~\ref{eq:mean_structure} as estimates of their standard deviations. This gives us a multivariate normal distribution for each subject's estimated parameter, 
\begin{equation}
\hat{\theta}_i \sim N(\theta_i, s_i^2).
\end{equation}
As our goal remains as being able to obtain reasonable estimates of the population curves for each group, it is necessary to estimate both the observed within-subject variability found in each of the $\{s_i^2\}$ terms, \textit{as well as} the between-subject variability present in $V_{g}$. For example, let $\theta^*_{ib}$ represent a bootstrapped sample for subject $i$ in bootstrap $b = 1, \dots, B$, where
\begin{equation}\label{eq:sub_boot_dist}
\theta^*_{ib} \sim N(\hat{\theta}_i, s_i^2),
\end{equation}
as was done in Step (2.) of the homogeneous bootstrapping algorithm. If we were to sample \textit{without replacement}, we would obtain a homogeneous mean value from the $b$th bootstrap for group $g$, $\theta^{(hom)}_{bg}$, where
\begin{equation}\label{eq:wo_rep_boot}
\theta^{(hom)}_{bg} = \frac{1}{n_g} \sum_{i=1}^{n_g} \theta^{*}_{ib}, \quad \theta^{(hom)}_{bg} \sim N \left( \mu_{g}, \frac{1}{n_g^2} \sum_{i=1}^{n_g} s_i^2 \right).
\end{equation}

Such an estimate captures the totality of the within-subject variability with each draw but fails to account for the variability in the group overall. For this reason, we sample the subjects \textit{with} replacement, creating the heterogeneous bootstrap mean $\theta_{bg}^{(het)}$, where again each $\theta_{ib}^*$ follows the distribution in Equation~\ref{eq:sub_boot_dist}, but the heterogeneous bootstrapped group mean now follows
\begin{equation}\label{eq:w_rep_boot}
\theta_{bg}^{(het)} \sim N \left( \mu_{g}, \frac{1}{n_g} V_{g} + \frac{1}{n_g^2} \sum s_i^2 \right).
\end{equation}

The estimated mean value remains unchanged, but the variability is now fully accounted for. We therefore present a modified version of the bootstrap which we call the \textit{heterogeneous bootstrap}, making the following changes to the original:  

\begin{enumerate}
\item In step (1), the specification of AR(1) structure is \textit{optional} and can be modified with arguments to functions in \xt{bdots}. Our simulations show that while failing to include it slightly inflates the type I error in the heterogeneous bootstrap when the data truly is auto-correlated, specifying an AR(1) structure can lead to overly conservative estimates when it is not.
\item In step (2), we sample subjects \textit{with replacement} and then for each drawn subject, randomly draw one bootstrap estimate for each of their model parameters based on the mean and standard errors derived from the \xt{gnls} estimate.
\end{enumerate}

Just as with the homogeneous bootstrap, these bootstrap estimates are used to create test statistics $T_t$ at each time point, written
\begin{equation}
T_t^{(b)} = \frac{(\overline{p}_{1t} - \overline{p}_{2t})}{\sqrt{s_{1t}^2 + s_{2t}^2}},
\end{equation}
where $\overline{p}_{gt}$ and $s_{gt}^2$ are mean and standard deviation estimates at each time point for groups $1$ and $2$, respectively. Finally, just as in \citet{oleson2017detecting}, one can use the autocorrelation of the $T_t^{(b)}$ statistics to create a modified $\alpha$ for controlling the FWER.

\subsection{Permutation Testing}

In addition to the heterogeneous bootstrap, we also introduce a permutation method for hypothesis testing. The permutation method proposed is analogous to a traditional permutation method, but with an added step mirroring that of the previous in capturing the within-subject variability. For a specified FWER of $\alpha$, the proposed permutation algorithm is as follows:

\begin{enumerate}
\vspace{-2mm}
\item For each subject, fit the nonlinear function with \textit{optional} AR(1) autocorrelation structure for model errors. Assuming large sample normality, the sampling distribution of each estimator can be approximated by a normal distribution with mean corresponding to the point estimate and standard deviation corresponding to the standard error
\item Using the mean parameter estimates derived in (1), find each subject's corresponding fixation curve. Within each group, use these to derive the mean and standard deviations of the population level curves at each time point, denoted $\overline{p}_{gt}$ and $s_{gt}^2$ for $g = 1,2$. Use these values to compute a permutation test statistic $T_t^{(p)}$ at each time point,
\begin{equation}
T_t^{(p)} = \frac{|\overline{p}_{1t} - \overline{p}_{2t}|}{\sqrt{s_{1t}^2 + s_{2t}^2}}.
\end{equation}
This will be our observed test statistic.
\item Repeat (2) $P$  additional times, each time shuffling the group membership between subjects. This time, when constructing each subject's corresponding fixation curve, draw a new set of parameter estimates using the distribution found in (1). Recalculate the test statistics $T_t^{(p)}$, retaining the maximum value from each permutation. This collection of $P$ statistics will serve as our null distribution which we denote $\widetilde{T}$. Let $\widetilde{T}_{\alpha}$ be the $1$ - $\alpha$ quantile of $\widetilde{T}$
\item Compare each of the observed $T_t^{(p)}$ with $\widetilde{T}_{\alpha}$. Areas where $T_t^{(p)} > \widetilde{T}_{\alpha}$ are designated significant. 
\end{enumerate}

\subsection{Paired Data}

Briefly, we attend to the issue of paired data for each of the discussed methods, as this is critical for a proper assessment of both FWER and power. Specifically, in a paired setting we note that our interest is in determining the distribution of paired differences rather than of the respective groups. 

For the homogeneous bootstrap, this is done by default: as each subject is sampled without replacement, we can be sure that each bootstrap estimate between groups contains the same subjects. For the heterogeneous bootstrap, this is done by ensuring that the same subjects sampled in each bootstrap for one group is matched identically with sampling subjects in the other. Lastly, in the case of permutation testing, paired data is addressed by ensuring that each permuted group contains one observation for each subjects, so that each paired subject in one permuted group has its corresponding observation in the other. 

\subsection{Modified Bonferonni Correction}

While the permutation method determines significance via comparison to an estimated null distribution, both the homogeneous and heterogeneous bootstrap construct estimates of the observed distribution against which the null hypothesis is tested. This results in a series of often highly correlated test statistics, raising concerns that are typically associated with multiple testing. 

The method for controlling FWER presented in \citet{oleson2017detecting} begins on the assumption that adjacent test statistics in a densely sampled time series will be highly correlated. That is, it will be assumed that the test statistics have null standard normal distribution $T_t \sim N(0,1)$, but that the sequence of statistics themselves unfolds with an AR(1) process
\begin{equation}
T_t = \rho  T_{t-1} + \epsilon_t,
\end{equation}
where $\epsilon_t \sim N(0, (1 - \rho^2))$. This gives the conditional distribution
\begin{equation}
T_t | T_{t-1} \sim N(\rho T_{t-1}, 1 - \rho^2),
\end{equation}
with the joint distribution of $T_t$ and $T_{t-1}$ being bivariate normal with mean $0$, a variance of $1$, and correlation $\rho$. From this, they show that for some $\alpha^*$, the true FWER $\alpha$ under the null hypothesis can be expressed as 
\begin{equation}\label{eq:oleson_adj}
1 - P \left( \bigcap_{t=1}^T I_t \right) = 1 - P(I_1) \prod_{t=2}^T P(I_t | I_{t-1}) = 1 - P(I_1)P(I_t|I_{t-1})^{T-1},
\end{equation}
where $I_t$ is the event that $|T_t| \leq z_{\left(1 - \frac{\alpha^*}{2} \right)}$. The correction is made by finding the nominal significance $\alpha^*$ that produces the desired FWER of $\alpha$. Of note here, no adjustment to the nominal $\alpha$ is needed when the tests are perfectly correlated. Conversely, when the tests are perfectly independent, Equation~\ref{eq:oleson_adj} reduces to the standard Bonferonni correction. This correction is implemented in \xt{bdots} and is also provided outside of the bootstrapping process with the function \xt{bdots::p\_adjust} using \xt{method = "oleson"}.

\section{FWER Simulations}

We now go about comparing the family-wise error rate of the three methods just described. In doing so, we will consider several conditions under which the observed subject data may have been generated or fit. This includes generating data with both a homogeneous and heterogeneous means assumption, generating data with and without autocorrelated errors, and fitting data with and without an AR(1) assumption. In addressing performance under paired and unpaired conditions, we have included two distinct but reasonable instances in which the data may be paired, which we will elaborate further on shortly. However, as paired data becomes difficult to define under the homogeneous means assumption (in which all subjects are, in a sense, ``paired"), we will omit these settings from our final simulations. In all, this gives us sixteen different arrangements which we will examine for their family-wise error rates using each of the three methods previously described.

\subsection{Data Generation}

Data was generated according to Equation~\ref{eq:mean_structure}, with the parametric function $f(t|\theta)$ belonging to the family of four-parameter logistic curves defined:
\begin{equation}\label{eq:logistic}
f(t | \theta) = \frac{p-b}{1 + \exp \left(\frac{4s}{\text{p}-b} (x - t) \right)} + b
\end{equation}
where $\theta = (p, b, s, x)$, the peak, baseline, slope, and crossover parameters, respectively.

\begin{figure}
    \centering
\includegraphics{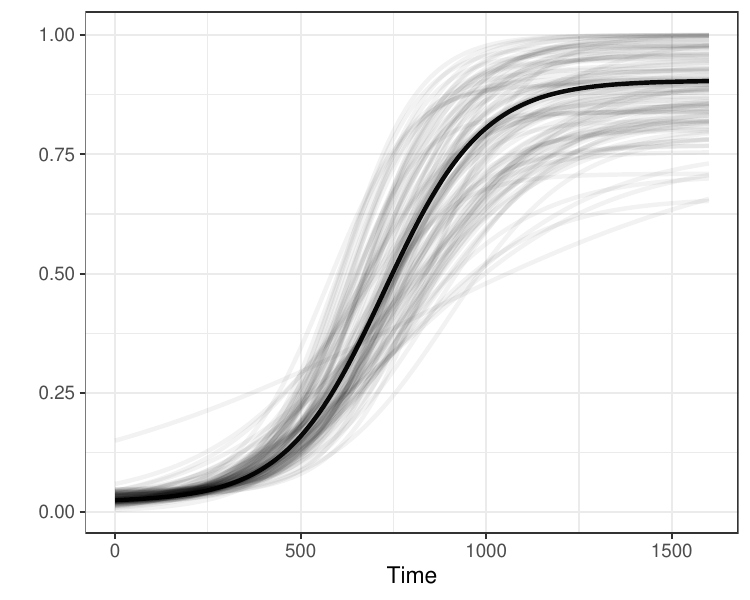}
    \caption{50 samples from the generating distribution of the four-parameter logistic in Equation~\ref{eq:logistic} used for testing FWER}
\label{fig:distribution_log}
\end{figure}

We further assume that each group drew subject-specific parameters from a multivariate normal distribution, with subject $i = 1, \dots, N$ in group $g = 1, \dots, G$ following the distribution in Equation~\ref{eq:theta_i_dist}. These parameters are then used to simulate empirical data according the the mean and error structures for each simulation.

\paragraph{Mean Structure} In all of the simulations presented, the distribution of parameters used in Equation~\ref{eq:theta_i_dist} was empirically determined from data on normal hearing subjects in the VWP \citep{FarrisTrimble2014}. Parameters used were those fit to fixations on the Target, following the functional form of Equation~\ref{eq:logistic}. A visual depiction of the distribution of these curves is given in Figure~\ref{fig:distribution_log}.

Under the homogeneous means assumption, we set $\theta_i = \theta_j$ for all subjects $i,j$, assuring that each of the subjects' observations is derived from the same mean structure, differing only in their observed error structure.

\paragraph{Error Structure} The error structure is of the form
\begin{equation}
e_{it} = \phi e_{i, t-1} + w_{it}, \quad w_{it} \sim N(0, \sigma)
\end{equation}
where the $w_{it}$ are iid with $\sigma = 0.025$. $\phi$ corresponds to an autocorrelation parameter and is set to $\phi = 0.8$ when the generated data is to be autocorrelated and set to $\phi = 0$ when we assume the errors are all independent and identically distributed. 

\paragraph{Paired Data} As we previously noted, paired data is only a sensible condition under the assumption of heterogeneous means, and we limit our consideration to that case. There are two methods that seem reasonable in the construction of paired data, and we employ both of them here. 

In considering the construction of paired data for subject $i$, the first method proceeds as follows: we begin by drawing parameters $\theta_i$ from Equation~\ref{eq:theta_i_dist}. Denote this $\theta_{i1}$ to indicate that this is the parameter estimate for subject $i$ in group $1$. We then simulate observed data according to Equation~\ref{eq:mean_structure}. To create the paired data, we then set $\theta_{i2} = \theta_{i1}$ and again simulate observed data according to Equation~\ref{eq:mean_structure}. Under this first method, the generating parameters between groups are \textit{identical}, with the only differences between the simulated data being that contributed by the error term.

In considering the construction of paired data for subject $i$, the first method proceeds as follows: under each experimental condition, $g$, set $\theta_{ig}$ so that $\theta_{i1} = \theta_{i2}$, where $\theta_{i1}$ is original drawn from Eq. \ref{eq:theta_i_dist}. We then simulate the observed data according to the mean struction in Eq. ~\ref{eq:mean_structure}. Under this method, the generating parameters for each individual are identical, with the only differences being those contributed by the error term.

In the second method of obtaining paired data, we proceed as in the first, except now letting
\begin{equation}
\theta_{i2} = \theta_{i1} + N(0, 0.05 \cdot  V_1).
\end{equation}
This adds a small amount of random noise between paired parameters, simulating the degree of variability that may normally be found between conditions, even when there is no true effect. Accommodating this phenomenon is relevant in situations in which the data gathering mechanism has imperfect reliability, as in the case of the VWP. It is also relevant because, as we will show, the homogeneous bootstrap is highly sensitive to these assumptions, with potentially disastrous results when these conditions are not precisely met. To avoid potential confusion, the results for each of these will be presented separately.

Each set of conditions generates two groups, with $n = 25$ subjects in each group, with time points $t = 0, 4, 8, \dots, 1600$ in each trial and with $100$ simulated trials for each subject. Columns in the tables indicate homogeneity of means assumption, whether or not an AR(1) error structure was used in constructing the data and if autocorrelation was specified in the fitting function. The last conditions help assess the impact of correctly or incorrectly identifying the type of error when conducting an analysis in \xt{bdots}. Finally, results will be separated by paired status. Each simulation was conducted 1000 times, with the proportion of simulations in which a significance difference was incorrectly identified used to determine the family-wise error rate.

\subsection{Results}

We consider the efficacy the methods under each of the simulation settings with an analysis of the family-wise error rate (FWER) and the median per-comparison error rate. The first of these details the proportion of simulations under each condition that marked \textit{at least} one time point as being significantly different between the two groups. This is critical is understanding each method's ability to correct adjust for the multiple testing problem associated with testing each of the observed time points. The results for the unpaired data are presented in Table~\ref{tab:fwer_unpaired}, while those for each of the paired variations are presented in Tables~\ref{tab:fwer_paired} and~\ref{tab:fwer_paired2}.

Complimenting the FWER estimate is an estimate of the median per-comparison rate. For each time point across each of the simulations, we computed the proportion of instances in which that time was incorrectly determined to have a significant difference. The median of these values across all time points is what is considered. This metric gives a sense of magnitude to the otherwise binary FWER; for example, a situation in which there was a high FWER and low median per-comparison rate would indicate that the type I error within a particular time series would be sporadic and impact limited regions. Large median per-comparison rates indicate that large swaths of a time series frequently sustain type I errors. The median per-comparison rates for unpaired simulations are presented in Table~\ref{tab:mpc_unpaired} and each of the paired simulations in Table~\ref{tab:mpc_paired1} and Table~\ref{tab:mpc_paired2}.

\subsubsection{FWER}

There are a few things of immediate note when considering the results of Table~\ref{tab:fwer_unpaired}. First, we see from the first two settings of the unpaired simulations that the FWER for the homogeneous bootstrap are consistent with those presented in \citet{oleson2017detecting}, confirming the importance of specifying the existence of autocorrelation in the \xt{bdots} fitting function when autocorrelated error is present. By contrast, this is far less of a concern when using the heterogeneous bootstrap or permutation testing, both of which maintain a FWER near the nominal alpha, regardless of whether or not the error structure was correctly identified. This continues to be true under the homogeneous mean assumption when the true error structure is not autocorrelated. 

The most striking results of this, however, appear when the data generation assumes a heterogeneous mean structure. While both the heterogeneous bootstrap and the permutation test maintain a FWER near the nominal alpha, the homogeneous bootstrap fails entirely, with a FWER $> 0.9$ in all cases.

\begin{table}[H]
\centering
\begin{tabular}{cccccc}
  \hline
  \multicolumn{1}{p{2cm}}{\centering Het. \\ Means} & \multicolumn{1}{p{2cm}}{\centering AR(1) \\ Error} & \multicolumn{1}{p{2cm}}{\centering AR(1) \\ Specified} &  \multicolumn{1}{p{2cm}}{\centering Hom. \\ Boot} &\multicolumn{1}{p{2cm}}{\centering Het. \\ Boot} & \multicolumn{1}{p{2cm}}{\centering Perm. } \\ 
  \hline
No & Yes & Yes & 0.09 & 0.00 & 0.06 \\ 
  No & Yes & No & 0.84 & 0.06 & 0.14 \\ 
  No & No & Yes & 0.12 & 0.01 & 0.08 \\ 
  No & No & No & 0.14 & 0.00 & 0.05 \\ 
  Yes & Yes & Yes & 0.94 & 0.05 & 0.05 \\ 
  Yes & Yes & No & 0.99 & 0.07 & 0.07 \\ 
  Yes & No & Yes & 1.00 & 0.08 & 0.05 \\ 
  Yes & No & No & 0.99 & 0.05 & 0.04 \\ 
   \hline
\end{tabular}
\caption{FWER for logistic function (unpaired)}
\label{tab:fwer_unpaired}
\end{table}

Table~\ref{tab:fwer_paired} gives the results under the construction assuming \textit{identical} parameters between groups. By construction, these results possess a homogenous mean structure and the FWER associated with the homogeneous bootstrap is comparable to that of the permutation test (though both with elevated FWER rates). Because these data are paired, the observed variability within each group has no effect on the distribution of paired differences. This is in stark contrast to the results presented in Table~\ref{tab:fwer_paired2}, where a small amount of variability was added to the paired set of parameters. What is most relevant here is that both the heterogeneous bootstrap as well as the permutation test are robust to these small differences in the paired setting, while the results associated with the homogeneous bootstrap are radically different. This serves to demonstrate how sensitive the homogeneous bootstrap is to such a rigid set of underlying assumptions.

\begin{table}[H]
\centering
\begin{tabular}{cccccc}
  \hline
  \multicolumn{1}{p{2cm}}{\centering Het. \\ Means} & \multicolumn{1}{p{2cm}}{\centering AR(1) \\ Error} & \multicolumn{1}{p{2cm}}{\centering AR(1) \\ Specified} &  \multicolumn{1}{p{2cm}}{\centering Hom. \\ Boot} &\multicolumn{1}{p{2cm}}{\centering Het. \\ Boot} & \multicolumn{1}{p{2cm}}{\centering Perm. } \\ 
  \hline
No & Yes & Yes & 0.10 & 0.00 & 0.12 \\ 
  No & Yes & No & 0.75 & 0.06 & 0.12 \\ 
  No & No & Yes & 0.12 & 0.00 & 0.11 \\ 
  No & No & No & 0.11 & 0.01 & 0.13 \\ 
   \hline
\end{tabular}
\caption{FWER for logistic function (paired, identical parameters)}
\label{tab:fwer_paired}
\end{table}

\begin{table}[H]
\centering
\begin{tabular}{cccccc}
  \hline
  \multicolumn{1}{p{2cm}}{\centering Het. \\ Means} & \multicolumn{1}{p{2cm}}{\centering AR(1) \\ Error} & \multicolumn{1}{p{2cm}}{\centering AR(1) \\ Specified} &  \multicolumn{1}{p{2cm}}{\centering Hom. \\ Boot} &\multicolumn{1}{p{2cm}}{\centering Het. \\ Boot} & \multicolumn{1}{p{2cm}}{\centering Perm. } \\ 
  \hline
Yes & Yes & Yes & 0.48 & 0.04 & 0.10 \\ 
  Yes & Yes & No & 0.93 & 0.07 & 0.12 \\ 
  Yes & No & Yes & 0.81 & 0.04 & 0.08 \\ 
  Yes & No & No & 0.81 & 0.07 & 0.09 \\
   \hline
\end{tabular}
\caption{FWER for logistic function (paired, added noise)}
\label{tab:fwer_paired2}
\end{table}

\subsubsection{Median per-comparison error rate}

We next consider the median per-comparison error rate, which offers some insight beyond what is provided by the FWER. In particular, consider the situation in which in Table~\ref{tab:mpc_unpaired}, in the fourth row we see a median per-comparison error rate of 0.00 for the homogeneous bootstrap, despite Table~\ref{tab:fwer_unpaired} indicating a FWER of 0.15. This is a consequence of the majority of the type I errors occurring in a relatively limited region. In contrast, the median per-comparison error rate of the homogeneous bootstrap under the assumption of heterogeneity suggests that the type I errors are widespread and not limited to any particular area. 

It is also worth commenting on the permutation test median per-comparison error rate in Table~\ref{tab:mpc_unpaired}; combined with a FWER near the nominal 0.05, that these values are not identically 0 suggests that errors are likely distributed across the entire range rather than limited to a small area.

\begin{table}[H]
\centering
\begin{tabular}{cccccc}
  \hline
  \multicolumn{1}{p{2cm}}{\centering Het. \\ Means} & \multicolumn{1}{p{2cm}}{\centering AR(1) \\ Error} & \multicolumn{1}{p{2cm}}{\centering AR(1) \\ Specified} &  \multicolumn{1}{p{2cm}}{\centering Hom. \\ Boot} &\multicolumn{1}{p{2cm}}{\centering Het. \\ Boot} & \multicolumn{1}{p{2cm}}{\centering Perm. } \\ 
  \hline
No & Yes & Yes & 0.02 & 0.00 & 0.01 \\ 
  No & Yes & No & 0.31 & 0.01 & 0.02 \\ 
  No & No & Yes & 0.01 & 0.00 & 0.01 \\ 
  No & No & No & 0.00 & 0.00 & 0.01 \\ 
  Yes & Yes & Yes & 0.59 & 0.01 & 0.01 \\ 
  Yes & Yes & No & 0.83 & 0.02 & 0.01 \\ 
  Yes & No & Yes & 0.84 & 0.02 & 0.01 \\ 
  Yes & No & No & 0.82 & 0.01 & 0.01 \\ 
   \hline
\end{tabular}
\caption{Median per-comparison error rate for unpaired data}
\label{tab:mpc_unpaired}
\end{table}

Regarding paired data, we see again a similar situation play out as we did with the FWER. Table~\ref{tab:mpc_paired1} treats the paired setting with identical parameters, and we see median per-comparison error rates consistent with the lower FWER from Table~\ref{tab:fwer_paired}. And again, Table~\ref{tab:mpc_paired2} demonstrates a more robust performance for both the heterogeneous bootstrap and permutation test.

\begin{table}[H]
\centering
\begin{tabular}{cccccc}
  \hline
  \multicolumn{1}{p{2cm}}{\centering Het. \\ Means} & \multicolumn{1}{p{2cm}}{\centering AR(1) \\ Error} & \multicolumn{1}{p{2cm}}{\centering AR(1) \\ Specified} &  \multicolumn{1}{p{2cm}}{\centering Hom. \\ Boot} &\multicolumn{1}{p{2cm}}{\centering Het. \\ Boot} & \multicolumn{1}{p{2cm}}{\centering Perm. } \\ 
  \hline
No & Yes & Yes & 0.03 & 0.00 & 0.02 \\ 
  No & Yes & No & 0.27 & 0.01 & 0.02 \\ 
  No & No & Yes & 0.01 & 0.00 & 0.02 \\ 
  No & No & No & 0.01 & 0.00 & 0.02 \\
   \hline
\end{tabular}
\caption{Median per-comparison error rate for paired data (identical parameters)}
\label{tab:mpc_paired1}
\end{table}

\begin{table}[H]
\centering
\begin{tabular}{cccccc}
  \hline
  \multicolumn{1}{p{2cm}}{\centering Het. \\ Means} & \multicolumn{1}{p{2cm}}{\centering AR(1) \\ Error} & \multicolumn{1}{p{2cm}}{\centering AR(1) \\ Specified} &  \multicolumn{1}{p{2cm}}{\centering Hom. \\ Boot} &\multicolumn{1}{p{2cm}}{\centering Het. \\ Boot} & \multicolumn{1}{p{2cm}}{\centering Perm. } \\ 
  \hline
Yes & Yes & Yes & 0.14 & 0.01 & 0.02 \\ 
  Yes & Yes & No & 0.46 & 0.01 & 0.03 \\ 
  Yes & No & Yes & 0.44 & 0.01 & 0.01 \\ 
  Yes & No & No & 0.41 & 0.02 & 0.02 \\ 
   \hline
\end{tabular}
\caption{Median per-comparison error rate for paired data (added noise)}
\label{tab:mpc_paired2}
\end{table}

\subsection{Discussion}

It was the control of the family-wise error rate in the context of densely sampled, highly correlated tests that first motivated the development of \xt{bdots}. While the results presented in this section indeed demonstrate control of the FWER under a strict set of assumptions, they also highlight the consequences when these conditions are not met. With the introduction of both the heterogeneous bootstrap and the permutation test, we have offered two alternatives that are robust to a wide variety of situations while maintaining performance similar to that of the homogeneous bootstrap in the best of cases. This is but one half of the question, however; in the following section, we present two separate simulations to determine if the robustness acquired is at the cost of significant power.

\section{Power Simulations}

In our assessment of power among the presented methods, we examine two distinct simulations, each associated with a different aspect of the question of identifying temporal differences. The first of these seeks to identify the relationship between power and effect size, which is done by comparing two simple piecewise linear functions. The goals of the second simulation, which is done in the context of VWP data, are twofold: first, to assess and verify the power of the new methods in paired and unpaired settings, and second, to investigate the relationship between power and effect size in light of between-subject variability.

\subsection{Piecewise Linear Power}

\begin{figure}[H]
    \centering
    \includegraphics{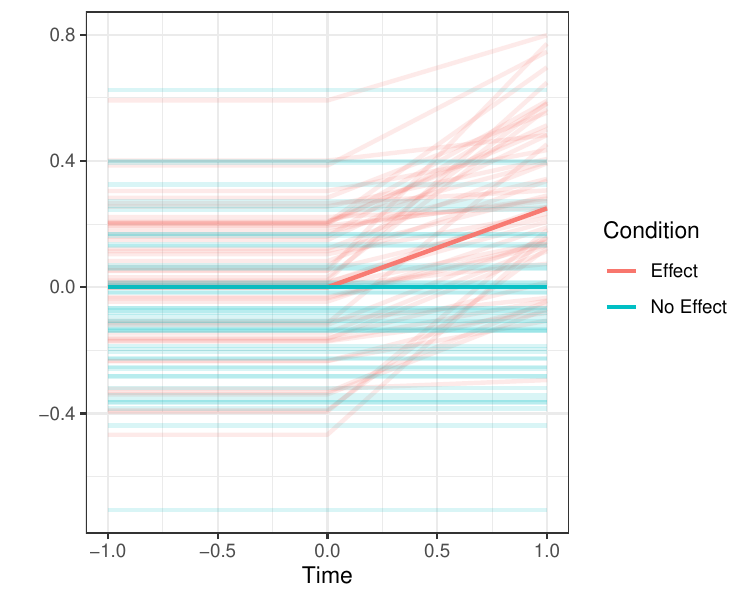}
    \caption{50 samples from the generating distributions of each group in Equation~\ref{eq:piecewise_form}. The distribution of values is the same for each group for all $t < 0$}
\label{fig:distribution_piece}
\end{figure}

To better understand the relationship between effect size and power, we have created a simple example consisting of two simulated experimental groups whose mean structure is a simple piecewise linear function defined on the interval $(-1, 1)$ as follows:
\begin{equation}\label{eq:piecewise_form}
y = \begin{cases}
b \quad &t < 0 \\
mx + b \quad &t \geq 0
\end{cases}
\end{equation}
The set of parameters drawn from each subject include a baseline parameter, $b$, as well as a slope parameter, $m$, which were each drawn from a univariate normal distribution. To distinguish experimental groups, one designated as being the ``Effect" group, other other being ``No Effect", we set the slope parameter of the ``No Effect" group to be identically $m = 0$, while the ``Effect" group had a slope parameter that was normally distributed with a mean value $\mu_m = 0.25$. Both groups drew their baseline parameter, $b$, from a normal distribution with mean $\mu_b = 0$. As a consequence of this, there should be no difference between groups when $x < 0$, and an effect size for the ``Effect" group of $mt$ for all $t \geq 0$. A visual depiction of each of these groups is given in Figure~\ref{fig:distribution_piece}.

For these simulations, we limited consideration to three possible scenarios: first, we assumed the conditions presented in \citet{oleson2017detecting}, assuming homogeneity between subject parameters and an AR(1) error structure, with the model fitting performed assuming autocorrelated errors. For the remaining scenarios, we assumed heterogeneity in the distribution of subject parameters, simulated with and without an AR(1) error structure. In both of these last two scenarios, we elected to \textit{not} fit the model assuming autocorrelated errors. This was for two reasons: first, simulations exploring the type I error rate suggested that models fit with the autocorrelation assumption tended to be conservative. Second, and given the results of the first, this makes setting the assumption of autocorrelation to FALSE in \xt{bdots} seem like a sensible default, and as such, it would be of interest to see how the model performs in cases in which there is autocorrelated error that is not accounted for. Simulation results for a number of other settings are handled in the appendix. 

\begin{figure}[H]
\centering.
\includegraphics{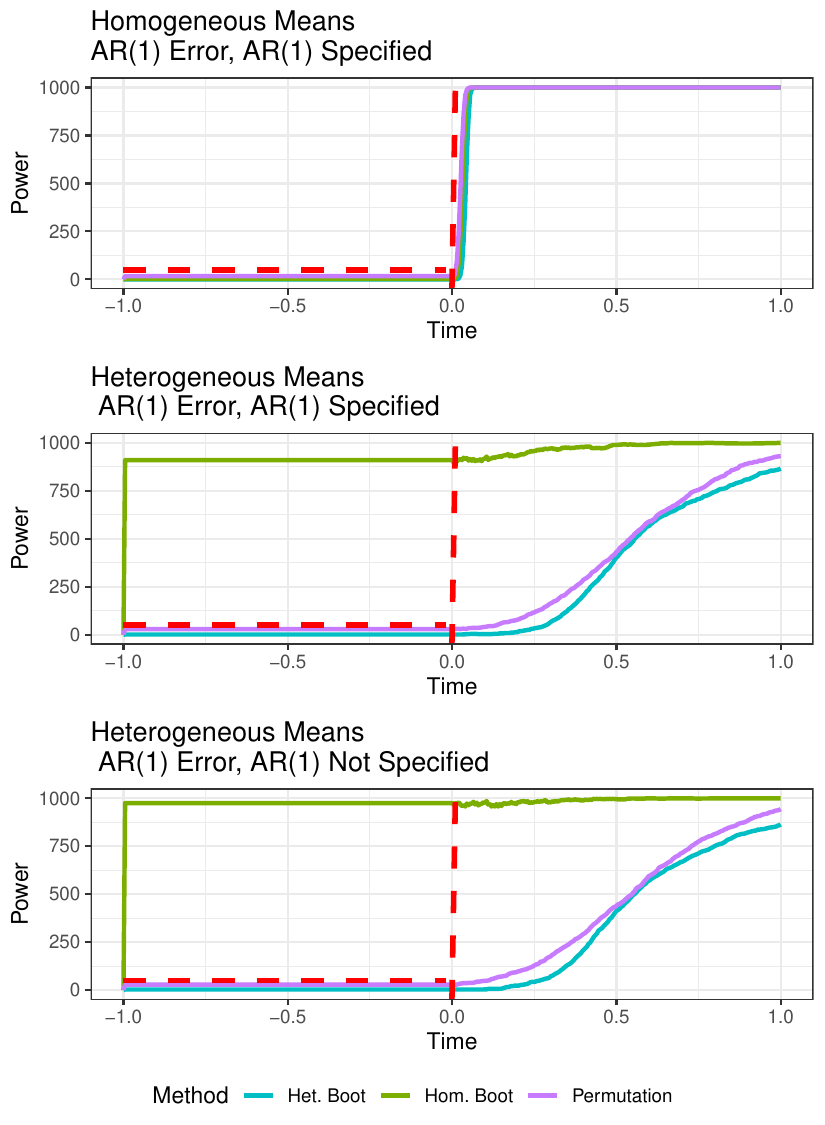}
\caption{Observed power of each of the methods at each time in (-1,1)}
\label{fig:time_power_plot}
\end{figure}

For each subject, parameters for their mean structure given in Equation~\ref{eq:piecewise_form} were drawn according to their group membership and fit using \xt{bdots} on the interval (-1,1). Time windows in which the groups differed were identified using the homogeneous bootstrap, heterogeneous bootstrap, and permutation testing. By including the interval (-1,0) in which the null hypothesis was true, we are able to mitigate the effects of over-zealous methods in determining power, and we present the results in the following way: any tests in which a difference was detected in (-1,0) was marked as having a type I error, and the proportion of simulations in which this occurred for each method is reported as the FWER in the column labeled $\alpha$. The next column, $\beta$, is the type II error rate, indicating the proportion of trials in which no differences were identified over the entire region. The last Greek-letter column is $1 - \beta - \alpha$, a modified power statistic indicating the proportion of tests in there were no errors across the interval. The remaining columns relate to this modified power column, giving a partial summary of the earliest onset time of detection. As a true difference occurs on the interval $t > 0$, smaller values indicate greater power in detecting differences. Finally, a plot giving the power at each time point is given in Figure~\ref{fig:time_power_plot}. This plot represents the true power, though note that it does not take into account the rate at which these regions were identified in conjunction with a type I error rate.

\subsubsection{Results}

The results of the power simulation are presented in Table~\ref{tab:power_methods}. We begin by considering the case in which we assumed a homogeneous mean structure with autocorrelated errors, matching the conditions in which the homogeneous bootstrap was first presented. Notably, we find that the permutation method demonstrates the greater power, with the median onset time just under that of the homogeneous bootstrap. This is at the expense of a larger FWER, though still below the nominal level. Alternatively, the heterogeneous bootstrap maintains a similar FWER as the homogeneous bootstrap at the cost of power. The remaining settings tell a similar story with the exception of the homogeneous bootstrap which continues to demonstrate unacceptable FWER under the heterogeneous means assumption. As to the effect of incorrectly specifying an AR(1) error structure when comparing the last two settings, note that there tends to be very little effect between the heterogeneous bootstrap and permutation, with neither performing consistently better or worse at any of the quantiles given.

\begin{table}[H]
\centering
\begin{tabular}{lcccccccc}
  \hline
Method & Heterogeneity & AR(1) & $\alpha$ & $\beta$ & 1 - $\alpha$ - $\beta$ & 1st Qu. & Median & 3rd Qu. \\ 
  \hline
Hom. Boot & No & Yes & 0.00 & 0.00 & 1.00 & 0.025 & 0.030 & 0.035 \\ 
  Het. Boot & No & Yes & 0.00 & 0.00 & 1.00 & 0.035 & 0.040 & 0.045 \\ 
  Perm & No & Yes & 0.03 & 0.00 & 0.97 & 0.020 & 0.025 & 0.030 \\
  \hline
  Hom. Boot & Yes & No & 0.95 & 0.00 & 0.05 & 0.005 & 0.008 & 0.010 \\ 
  Het. Boot & Yes & No & 0.00 & 0.01 & 0.98 & 0.260 & 0.330 & 0.480 \\ 
  Perm & Yes & No & 0.04 & 0.00 & 0.95 & 0.245 & 0.325 & 0.452 \\
  \hline
  Hom. Boot & Yes & Yes & 0.94 & 0.00 & 0.06 & 0.005 & 0.013 & 0.015 \\ 
  Het. Boot & Yes & Yes & 0.01 & 0.01 & 0.98 & 0.270 & 0.370 & 0.465 \\ 
  Perm & Yes & Yes & 0.04 & 0.00 & 0.96 & 0.245 & 0.365 & 0.440 \\ 
   \hline
\end{tabular}
\caption{Summary of FWER and power across methods} 
\label{tab:power_methods}
\end{table}

The results in Table~\ref{tab:type_2_summary} are a summary of all of the methods found by taking the mean of each of the results presented. This is intended to interrogate the performance of each of these methods when underlying assumptions are unknown or unspecified. We find a robust performance for each of the new methods presented, maintaining a reasonable relationship between FWER and power. The metrics associated with homogeneous bootstrap are perhaps a bit misleading here as they appear to demonstrate exceptional power, though at the cost of unacceptable type I error.

\begin{table}[H]
\centering
\begin{tabular}{lcccccc}
  \hline
Method & $\alpha$ & $\beta$ & $1 - \alpha - \beta$ & 1st Qu. & Median & 3rd Qu. \\ 
  \hline
Hom. Bootstrap & 0.772 & 0.000 & 0.228 & 0.011 & 0.016 & 0.021 \\ 
  Het. Bootstrap & 0.002 & 0.097 & 0.901 & 0.328 & 0.419 & 0.546 \\ 
  Permutation & 0.027 & 0.038 & 0.935 & 0.294 & 0.424 & 0.556 \\ 
   \hline
\end{tabular}
\caption{Summary of methods for Type II error} 
\label{tab:type_2_summary}
\end{table}

\subsection{Logistic Shift Power}

The final set of simulations we consider seek to address concerns related to power in a paired setting, as well as to investigate the relationship between effect size and variability. We proceed with simulation settings similar to those in our investigation of the type I error in that each of the two groups under consideration draw from an empirical distribution of parameters associated with the four parameter logistic function. Unlike the situation in determining the FWER, here differences are introduced between groups by changing the crossover parameter, indicating the inflection point of the function. To simulate various effect sizes, we investigate changes in crossover of $50$ and $150$. Further, to determine the effect of between subject variability on the identification of differences, we have set the standard deviation of the crossover parameter from the empirical distribution to take values of either $60$ or $120$. Thus, the settings we will be presenting include small and large changes in light of small or large variability. Data was generated with iid errors and fit without specifying AR(1) correlation in \xt{bdots}. Finally, each of these settings will be run as paired and unpaired data, with parameters for the paired data being identical between groups (i.e., no added variability).

\begin{figure}[t]
\centering.
\includegraphics{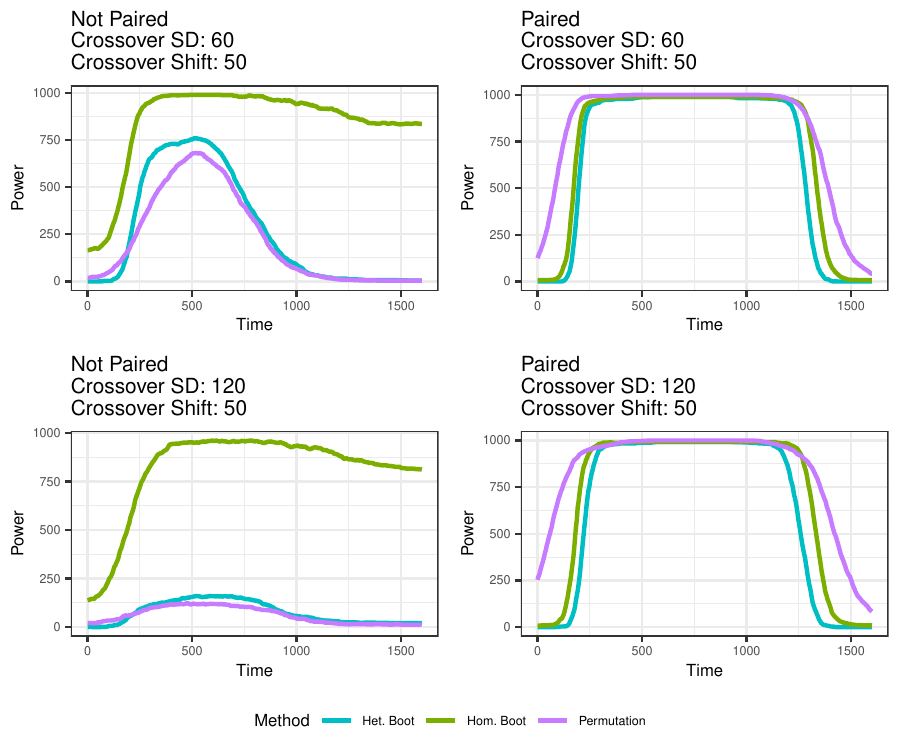}
\caption{Observed power following a small shift in the crossover parameter}
\label{fig:log_shift_1}
\end{figure}

Consider first, for example, the results presented in Figure~\ref{fig:log_shift_1}, each of which demonstrate the observed power at each time point when the crossover parameter between group was shifted by $50$. Predictably, we see greater power in the unpaired setting when the standard deviation of the crossover parameter is smaller, relative to the shift. By contrast, in the paired setting we observe little difference in power as a function of parameter variability. This is also to be expected as here we are concerned with the variability of the \textit{difference} rather than of each of the groups, and these results confirm that this is indeed the case.

\begin{figure}[t]
\centering.
\includegraphics{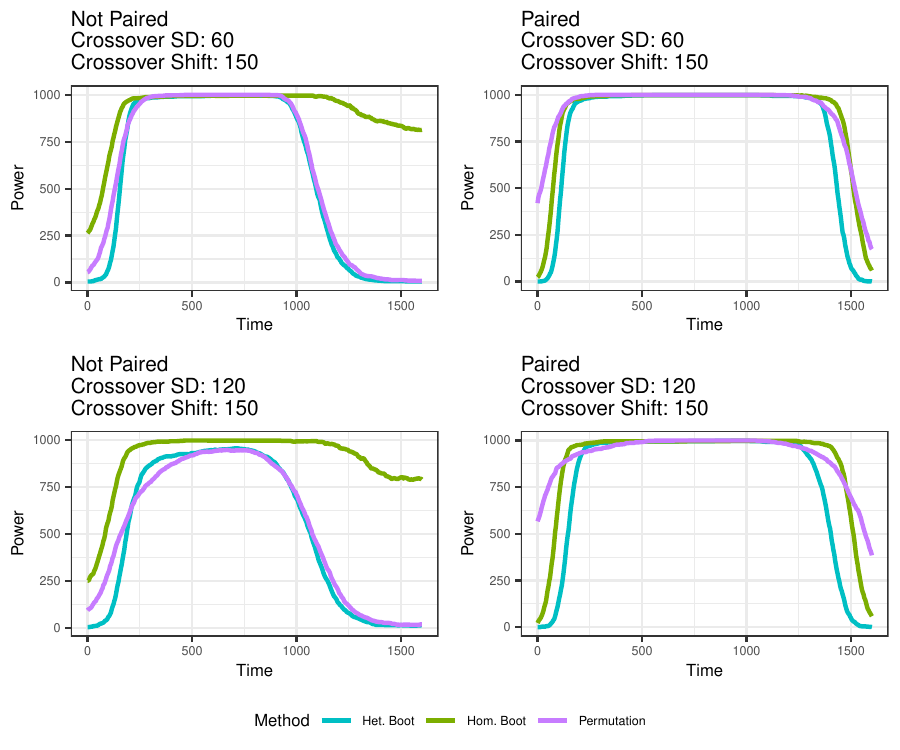}
\caption{Observed power following a large shift in the crossover parameter}
\label{fig:log_shift_2}
\end{figure}

A similar set of conclusions can be seen by considering the observed power presented in Figure~\ref{fig:log_shift_2}, where we observe the results when the crossover parameter is shifted by $150$. Similar to the prior case, we see greater power associated with the smaller standard deviation in the unpaired case, with power in the paired case performing similarly. And finally, by comparing the results between Figures~\ref{fig:log_shift_1} and~\ref{fig:log_shift_2} we confirm in all cases that a greater effect size is associated with greater power. 

\section{Discussion}

We set out both to interrogate the validity of the homogeneous bootstrap assumptions and to propose two alternative methods that would be more robust under a greater variety of assumptions. In doing so, we demonstrated conclusively the utility of the heterogeneous bootstrap and permutation tests while also highlighting a major shortcoming of the original. It's worth noting, however, that the FWER adjustment proposed in \cite{oleson2017detecting} is still valid, if not slightly conservative, and with power similar to that of the permutation method.

There are several limitations of the current paper that are worthy of further investigation. First, limited consideration was given to the effect of sample density on the observed type I error rate or power. As the fitting function in \xt{bdots} simply returns a set of parameters, one could conceivably perform any of the methods presented on any arbitrary collection of points, whether or not any data were observed there. This extends itself to the condition in which subjects were sampled at heterogeneous time points, as may be the case in many clinical settings. What impact this may have on how to best handle these cases remains open for exploration. It is also worth investigating in greater detail what impact the re-drawing of subject specific parameters from their respective distributions has on both the FWER and power, as in several of the simulations the observed FWER was much lower than the nominal level. Particularly in the case of the permutation method which is \textit{not} seeking to estimate the group distributions, it may be worthwhile to see if a favorable trade can be made to increase the resulting power.

We conclude by noting that \xt{bdots} is now equipped with two methods to effectively control the FWER when assessing the differences in time series under a greater set of underlying assumptions, including those involving the presence of highly correlated test statistics. Further, both methods presented are robust to misspecification of the error structure while maintaining an acceptable FWER and adequate power.

\bibliography{mybib}

\appendix

\input{append.tex}

\end{document}

%% file: append.tex
\section{Full Piecewise Power Simulations}

Here we present the full collection of the linear piecewise power simulations, which, in addition to those given in Table~\ref{tab:power_methods} includes cases for heterogeneous means where autocorrelation is specified when fitting with \xt{bdots}. This is indicated in the ``AR(1) Specified" column. Notably from this table, we see that in the case in which the true errors are iid, there is no measurable effect on power when an autocorrelated structure is incorrectly specified.  This is similar to the opposite situation, in which the true error does have an AR(1) structure. In this case, we observe a marginal benefit to correctly specifying an AR(1) structure. This may in fact make retaining the AR(1) assumption a reasonable default in the \xt{bdots} package.

\begin{figure}
\centering
\includegraphics[scale=1]{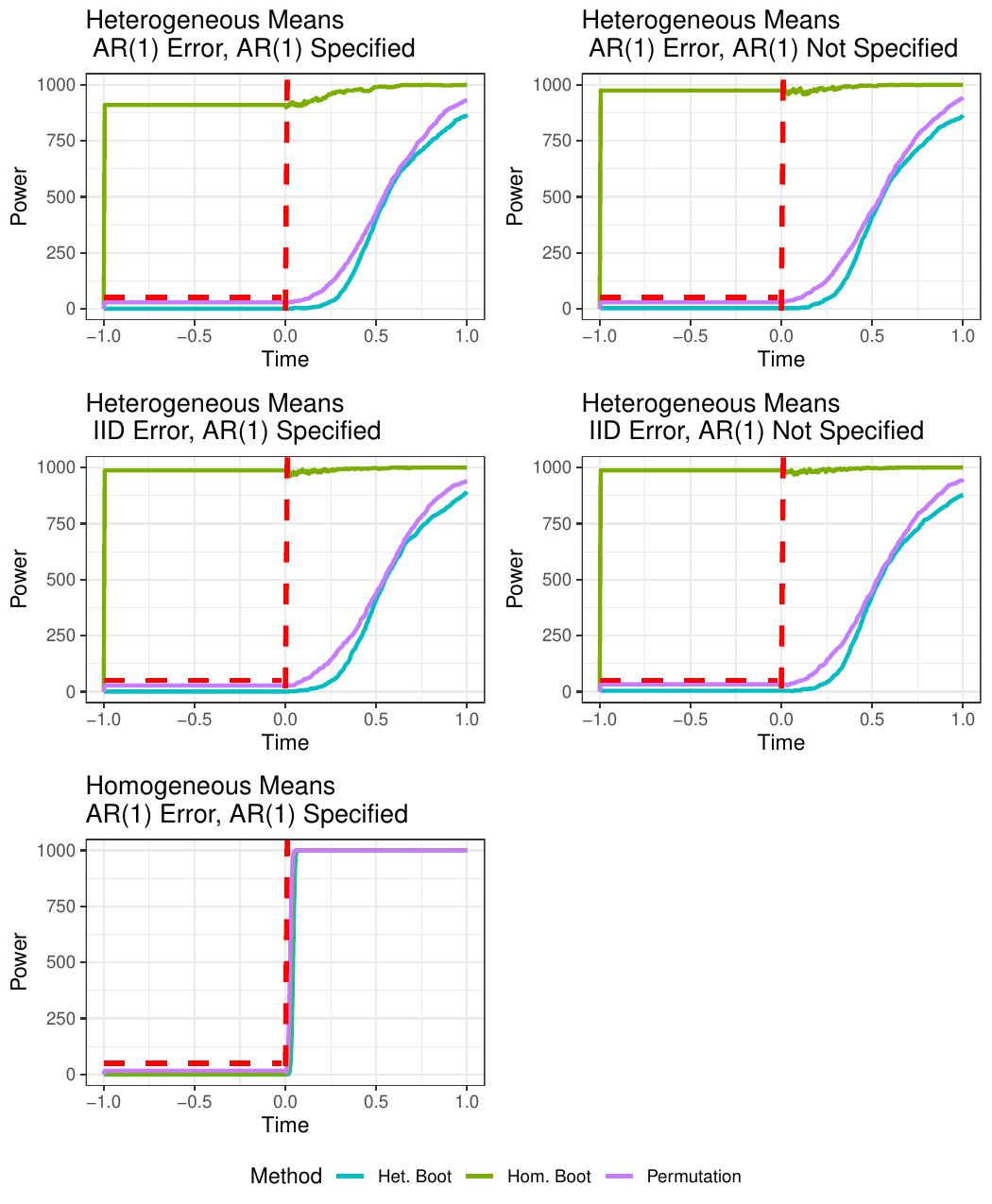}
\caption{Power plots in time for each of the simulation settings. Note that in the heterogeneous means case, there is little difference when AR(1) is incorrectly specified}
\label{fig:time_power_plot_full}
\end{figure}

\begin{landscape}
\begin{table}[ht]
\centering
\begin{tabular}{lccccccccc}
  \hline
Method & Heterogeneity & AR(1) Error & AR(1) Specified & $\alpha$ & $\beta$ & 1 - $\alpha$ - $\beta$ & 1st Qu. & Median & 3rd Qu.  \\ 
  \hline
Hom. Boot & No & Yes & Yes & 0.00 & 0.00 & 1.00 & 0.025 & 0.030 & 0.035 \\ 
  Het. Boot & No & Yes & Yes & 0.00 & 0.00 & 1.00 & 0.035 & 0.040 & 0.045 \\ 
  Perm & No & Yes & Yes & 0.03 & 0.00 & 0.97 & 0.020 & 0.025 & 0.030 \\ \hline
  Hom. Boot & Yes & No & No & 0.95 & 0.00 & 0.05 & 0.005 & 0.008 & 0.010 \\ 
  Het. Boot & Yes & No & No & 0.00 & 0.01 & 0.98 & 0.260 & 0.330 & 0.480 \\ 
  Perm & Yes & No & No & 0.04 & 0.00 & 0.95 & 0.245 & 0.325 & 0.452 \\ \hline
  Hom. Boot & Yes & No & Yes & 0.98 & 0.00 & 0.02 & 0.005 & 0.008 & 0.010 \\ 
  Het. Boot & Yes & No & Yes & 0.00 & 0.01 & 0.99 & 0.261 & 0.350 & 0.475 \\ 
  Perm & Yes & No & Yes & 0.04 & 0.00 & 0.96 & 0.225 & 0.335 & 0.440 \\ \hline
  Hom. Boot & Yes & Yes & No & 0.94 & 0.00 & 0.06 & 0.005 & 0.013 & 0.015 \\ 
  Het. Boot & Yes & Yes & No & 0.01 & 0.01 & 0.98 & 0.270 & 0.370 & 0.465 \\ 
  Perm & Yes & Yes & No & 0.04 & 0.00 & 0.96 & 0.245 & 0.365 & 0.440 \\ \hline
  Hom. Boot & Yes & Yes & Yes & 0.83 & 0.00 & 0.17 & 0.021 & 0.032 & 0.040 \\ 
  Het. Boot & Yes & Yes & Yes & 0.00 & 0.01 & 0.98 & 0.250 & 0.330 & 0.450 \\ 
  Perm & Yes & Yes & Yes & 0.03 & 0.00 & 0.97 & 0.223 & 0.335 & 0.428 \\ 
   \hline
\end{tabular}
\caption{Power for full piecewise linear simulation} 
\label{tab:power_methods_full}
\end{table}
\end{landscape}

%% file: method.bbl
\begin{thebibliography}{5}
\providecommand{\natexlab}[1]{#1}
\providecommand{\url}[1]{\texttt{#1}}
\expandafter\ifx\csname urlstyle\endcsname\relax
  \providecommand{\doi}[1]{doi: #1}\else
  \providecommand{\doi}{doi: \begingroup \urlstyle{rm}\Url}\fi

\bibitem[Nolte and Breheny(2025)]{nolte2025}
Collin Nolte and Patrick Breheny.
\newblock Reintroduction of the bdots r package: Methodological and syntactical
  changes to the bootstrapped differences of time series.
\newblock 2025.

\bibitem[Oleson et~al.(2017)Oleson, Cavanaugh, McMurray, and
  Brown]{oleson2017detecting}
Jacob~J Oleson, Joseph~E Cavanaugh, Bob McMurray, and Grant Brown.
\newblock Detecting time-specific differences between temporal nonlinear
  curves: Analyzing data from the visual world paradigm.
\newblock \emph{Statistical methods in medical research}, 26\penalty0
  (6):\penalty0 2708--2725, 2017.

\bibitem[Seedorff et~al.(2018)Seedorff, Oleson, and
  McMurray]{seedorff2018bdots}
Michael Seedorff, Jacob Oleson, and Bob McMurray.
\newblock Detecting when timeseries differ: Using the bootstrapped differences
  of timeseries (bdots) to analyze visual world paradigm data (and more).
\newblock \emph{Journal of memory and language}, 102:\penalty0 55--67, 2018.

\bibitem[Nolte et~al.(2022)Nolte, Seedorff, Oleson, Brown, Cavanaugh, and
  McMurray]{bdots2025}
Collin Nolte, Michael Seedorff, Jacob Oleson, Grant Brown, Joseph Cavanaugh,
  and Bob McMurray.
\newblock \emph{bdots: Bootstrapped Differences of Time Series}, 2022.
\newblock URL \url{https://github.com/collinn/bdots}.
\newblock R package version 2.0.0.

\bibitem[Farris-Trimble et~al.(2014)Farris-Trimble, McMurray, Cigrand, and
  Tomblin]{FarrisTrimble2014}
Ashley Farris-Trimble, Bob McMurray, Nicole Cigrand, and J.~Bruce Tomblin.
\newblock The process of spoken word recognition in the face of signal
  degradation.
\newblock \emph{Journal of Experimental Psychology: Human Perception and
  Performance}, 40\penalty0 (1):\penalty0 308--327, feb 2014.
\newblock \doi{10.1037/a0034353}.

\end{thebibliography}
